%% file: root.tex
\documentclass[letterpaper, 10 pt, conference]{ieeeconf}  % Comment this line out if you need a4paper
\IEEEoverridecommandlockouts                              % This command is only needed if 
 \pdfoutput=1 
\usepackage{cite}
\usepackage{graphics} % for pdf, bitmapped graphics files
\usepackage{epsfig} % for postscript graphics files
\usepackage{amsmath} % assumes amsmath package installed
\usepackage{amssymb}  % assumes amsmath package installed
\usepackage[english]{babel}
\usepackage{algorithm}
\usepackage{algpseudocode}
\usepackage{graphicx}
\usepackage[caption=false]{subfig}
\usepackage{lipsum} % for mock text

\newtheorem{remark}{Remark}
\newtheorem{proposition}{Proposition}

\usepackage{tabularx} % table allows line-break
\usepackage{fourier} 
\usepackage{array}
\usepackage{makecell}

\newcommand{\set}[1]{\left\{#1\right\}}
\newcommand{\bs}{\boldsymbol}

\newcommand{\norm}[1]{\left\Vert#1\right\Vert}

\title{\LARGE \bf
Numerical Discretization Methods \\ for the Extended Linear Quadratic Control Problem
}
 
\author{Zhanhao Zhang, Jan Lorenz Svensen, Morten Wahlgreen Kaysfeld, \\ 
Anders Hilmar Damm Christensen,  Steen Hørsholt, 
John Bagterp Jørgensen% <-this % stops a space
}

\begin{document}

\maketitle
\thispagestyle{empty}
\pagestyle{empty}

%%%%%%%%%%%%%%%%%%%%%%%%%%%%%%%%%%%%%%%%%%%%%%%%%%%%%%%%%%%%%%%%%%%%%%%%%%%%%%%%
\input{tex/abstract}

%%%%%%%%%%%%%%%%%%%%%%%%%%%%%%%%%%%%%%%%%%%%%%%%%%%%%%%%%%%%%%%%%%%%%%%%%%%%%%%%

%\input{tex/Introduction}
% \input{tex/IntroductionJBJO}
\input{tex/IntroductionZhaz}

% \input{tex/IntroductionDistribution_NoAppedix}
%\input{tex/DiscretizationProblems}
\input{tex/NumericalMethodsZHAZ}

\input{tex/NumericalExperiments}
\input{tex/Conclusion}
%\input{tex/ExtendedLQ}

% \newpage
\section{APPENDIX}
\input{tex/AppendixProofs}

\addtolength{\textheight}{-12cm}   % This command serves to balance the column lengths

%%%%%%%%%%%%%%%%%%%%%%%%%%%%%%%%%%%%%%%%%%%%%%%%%%%%%%%%%%%%%%%%%%%%%%%%%%%%%%%%

%%%%%%%%%%%%%%%%%%%%%%%%%%%%%%%%%%%%%%%%%%%%%%%%%%%%%%%%%%%%%%%%%%%%%%%%%%%%%%%%

%%%%%%%%%%%%%%%%%%%%%%%%%%%%%%%%%%%%%%%%%%%%%%%%%%%%%%%%%%%%%%%%%%%%%%%%%%%%%%%%

% \input{JanCopy}
% \input{tex/AppendixDistribution}

% \section*{ACKNOWLEDGMENT}
% XX

\bibliographystyle{IEEEtranS}
\bibliography{reference}

\end{document}

%% file: tex/abstract.tex
\begin{abstract}
In this study, we introduce numerical methods for discretizing continuous-time linear-quadratic optimal control problems (LQ-OCPs). The discretization of continuous-time LQ-OCPs is formulated into differential equation systems, and we can obtain the discrete equivalent by solving these systems. We present the ordinary differential equation (ODE), matrix exponential, and a novel step-doubling method for the discretization of LQ-OCPs. Utilizing Euler-Maruyama discretization with a fine step, we reformulate the costs of continuous-time stochastic LQ-OCPs into a quadratic form, and show that the stochastic cost follows the $\chi^2$ distribution. In the numerical experiment, we test and compare the proposed numerical methods. The results ensure that the discrete-time LQ-OCP derived using the proposed numerical methods is equivalent to the original problem.
\end{abstract}

%% file: tex/IntroductionZhaz.tex
\section{Introduction}
\label{sec:Intro}
In optimal control theory, linear-quadratic (LQ) optimal control problems are considered the fundamental but core problem. They involve a quadratic cost function that needs to be optimized, and the controlled system is linear. Due to the simplicity and analytical solvability of LQ-OCPs, it has widespread practical applications in numerous fields, including engineering, aerospace, biology and economics~\cite{BrysonAE1975AppliedOC, conway2010spacecraft, lenhart2007optimal, ljungqvist2004recursive}. Implementing continuous-time LQ-OCPs in real-world scenarios might be infeasible, primarily because many practical applications operate on digital platforms in digital form. Consequently, discretization techniques become imperative in practice.

There is rich research on discretization and numerical solution methods for optimal control problems~\cite{Cullum1969DiscreteApproximationstoContinuoustimeOCP,Goebel2007LQRwithControlConstraint,BAGTERPJORGENSEN2012187,Dontchev2001TheEulerApproximation, Han2010ConvergenceofDiscretetime, Martin2015ConvergenceResults, Alt2013ApproximationsofLQ}. However, this is not the case for how to obtain the discrete-time equivalent from a continuous-time LQ-OCP. Based on existing literature, there are two popular ways for LQ discretization: 1) design a continuous-time LQ-OCP and subsequently derive its discrete-time equivalent, 2) initially discretizing the continuous-time system, followed by designing a discrete-time LQ-OCP with the discrete-time system. Compared with the latter, the first method provides a better approximation, which eventually leads to the solution of the original problem. 

The continuous-time LQ-OCP can be converted into discrete-time by assuming zero-order hold (ZOH) on the state vector. However, this approximation is a crude approximation that is often inaccurate for large sample times~\cite{Hendricks2008LCD}. In~\cite{Karl1970IntroToStoContTheory, åström2011computer}, the cost equivalent is obtained by extending the continuous-time LQ-OCP's cost function. It leads to the analytic expressions of the desired equivalent, discrete weighting matrices. Further, a matrix exponential method is introduced for calculating the cost equivalent~\cite{franklin1990DigitalControl,NewStepDoubling2010Nigham, ExponentialIntegrators2011Higham, Moler2003NineteenDubiousWays25YearsLater, vanLoan1978MatrixExponential}. Modelling sampled-data systems with traditional approaches has fundamental difficulties, which can be resolved using incremental models. Incremental models provide a seamless connection
between continuous- and discrete-time systems, and they can be implemented for optimal filtering and control~\cite{Goodwin2014SampledDataModelsForLinear,Goodwin2013SamplingandSampledDataModels,Middleton1990DigitalCA,Goodwin1988ConnectionBetweenContinuous}. In addition, LQ-OCPs are associated with advanced control algorithms, such as model predictive control (MPC). Prediction-Error-Methods for identification of these models exist \cite{Jorgensen:Jorgensen:PEM:ACC2007,Jorgensen:Jorgensen:PEMforMPC:ACC2007,Jorgensen:Jorgensen:PEMforMPC:ECC2007,john2004MHEstimationAndControl}. Also efficient numerical methods for the solution of extended LQ-OCPs exists, e.g. structure exploiting algorithms that use a Riccati recursion~\cite{BAGTERPJORGENSEN2012187, John2013EfficientImplementationofRiccati,john2004MHEstimationAndControl}. LQ-OCPs become hard to solve and analyze when the controlled systems are stochastic. The analytic expression of the stochastic cost and its expectation can be calculated using It$\hat o$ calculus~\cite{Karl1970IntroToStoContTheory, åström2011computer}. Many stochastic LQ-OCPs aim to optimize the mean value of their costs, which may not be true for some scenarios, e.g., conditional Value-at-Risk optimization problems (CVaR optimization)~\cite{Pardalos2001StochasticOptimization,Rockafellar2001Conditionalvalueatrisk,Uryasev2000Conditionalvalueatrisk,CAPOLEI2015214,steen2019OperationOptimization}. Therefore, it is critical to investigate the cost function distribution of stochastic LQ-OCPs.

The key problem that we address in this paper: 
\begin{itemize}
    \item [1.] Formulation of differential equation systems for LQ discretization
    \item [2.] Numerical methods for solving the resulting systems of differential equations
    \item [3.] Distribution of stochastic cost functions
\end{itemize}
In Section~\ref{sec:LinearQuadraticOptimalControlProblems}, we introduce deterministic and stochastic LQ-OCPs and propose differential equation systems for LQ discretization. For stochastic LQ-OCPs, we reformulate their cost function and describe the distribution of the stochastic cost. Section~\ref{sec:NumericalMethods} introduces three numerical methods for solving proposed differential equation systems.  Section~\ref{sec:NumericalExperiments} presents a numerical experiment comparing the proposed numerical methods, and conclusions are given in Section~\ref{sec:Conclusion}.

\section{Linear-Quadratic Optimal Control Problems}
\label{sec:LinearQuadraticOptimalControlProblems}
In this section, we introduce deterministic and stochastic LQ-OCPs and describe the differential equation systems for LQ discretization.
\subsection{Deterministic linear-quadratic optimal control problem}
Consider the deterministic LQ-OCP
\begin{subequations}
\label{eq:Deterministic:ContinuousTime:LinearQuadraticOCP}
\begin{alignat}{5}
& \min_{x,u,z,\tilde z} \quad &&\phi = \int_{t_0}^{t_0+T} l_c(\tilde z(t)) dt \\
& s.t. && x(t_0) = \hat x_0, \\
& && u(t) = u_k, \quad &&t_k \leq t < t_{k+1}, \, k \in \mathcal{N}, \\
& && \dot x(t) = A_c x(t) + B_c u(t), \quad && t_0 \leq t < t_0+T, \\
& && z(t) = C_c x(t) + D_c u(t), \, && t_0 \leq t < t_0+T, \\
& && \bar z(t) = \bar z_k,  &&t_k \leq t < t_{k+1}, \, k \in \mathcal{N}, \\
& && \tilde z(t) = z(t) - \bar z(t), && t_0 \leq t < t_0+T,
\end{alignat}
\end{subequations}
with the stage cost function
\begin{equation}
\begin{split}
    l_c(\tilde z(t)) &= \frac{1}{2} \norm{ W_z \tilde z(t)  }_2^2
    = \frac{1}{2} \tilde z(t) ' Q_{c} \tilde z(t), 
\end{split}
\label{eq:det-stagecost}
\end{equation}
where $Q_{c} = W_z' W_z$ is a semi-positive definite matrix. This problem is in continuous-time with decision variables $x(t),u(t),z(t)$, and $\Tilde{z}(t)$. The control horizon $T = N T_s$ with sampling time $T_s$ and $N \in \mathbb{Z}^+$, and $\mathcal{N}=0,1,\ldots, N-1$. We assume piecewise constant inputs, $u(t) = u_k$ and target variables $\bar z(t) = \bar{z}_k$ for $t_k \leq t < t_{k+1}$. 

\begin{remark}
Note that the case $\bar{z}(t) = [\bar{x}(t); \, \bar{u}(t)]$, $C_c = \begin{bmatrix} I \\ 0 \end{bmatrix}$, $D_{c} = \begin{bmatrix} 0  \\ I \end{bmatrix}$, and $Q_{c} = \begin{bmatrix} Q_{c,xx} & 0 \\ 0 & Q_{c,uu}\end{bmatrix}$ corresponds to
\begin{equation}
\begin{split}
    l_c(\tilde z(t)) &= \frac{1}{2} \left[x(t) - \bar x(t)\right]' Q_{c,xx} \left[ x(t) - \bar x(t) \right] 
    \\ & \qquad + \frac{1}{2} \left[ u(t) - \bar u(t) \right]' Q_{c,uu} \left[u(t)-\bar u(t) \right].
\end{split}
\end{equation}
{\flushright \hfill $\blacksquare$}
\end{remark}
The corresponding discrete-time LQ-OCP is
\begin{subequations}
\label{eq:Deterministic:DiscreteTime:LinearQuadraticOCP}
\begin{alignat}{5}
& \min_{ x,u} \quad &&\phi = \sum_{k\in \mathcal{N}} l_k(x_k,u_k)  \\
& s.t. && x_0 = \hat x_0, \\
& && x_{k+1} = A x_k + B u_k, \quad && k \in \mathcal{N},
\end{alignat}    
\end{subequations}
with the stage costs
\begin{equation}
    l_k(x_k,u_k) = \frac{1}{2} \begin{bmatrix} x_k \\ u_k \end{bmatrix}' Q \begin{bmatrix} x_k \\ u_k \end{bmatrix} + q_k' \begin{bmatrix} x_k \\ u_k \end{bmatrix} + \rho_k, \quad k \in \mathcal{N},
\label{eq:deterministic-stageCost}
\end{equation}
where the coefficient in the affine term and the constant term are
\begin{equation}
    q_k =  M \bar{z}_k, \quad  \rho_k =  \int_{t_k}^{t_{k+1}} l_c(\bar{z}_k) dt = l_c(\bar z_k) T_s, \quad k \in \mathcal{N}.
\label{eq:qkandrho_k}
\end{equation}
% and the constant term is
% \begin{equation}
%     \rho_k =  \int_{t_k}^{t_{k+1}} l_c(\bar{z}_k) dt = l_c(\bar z_k) T_s, \quad k \in \mathcal{N}
% \end{equation}
% The key problem that we address in this paper is 1) formulation of differential equation systems that can be used to compute $(A,B,Q,M)$ such that \eqref{eq:Deterministic:DiscreteTime:LinearQuadraticOCP} is equivalent to \eqref{eq:Deterministic:ContinuousTime:LinearQuadraticOCP}, and 2) numerical methods for solving the resulting systems of differential equations.

\begin{proposition}[Discretization of the deterministic LQ-OCP] The system of differential equations
\label{prop:DiscretizationoftheDterministicLQOCP}
\begin{subequations}
\label{eq:DeterministicLQ:MatrixODEsystem}
\begin{alignat}{3}
    \dot A(t) &= A_c A(t), \qquad && A(0) = I, \\
    \dot B(t) &= A(t) B_c, \qquad && B(0) = 0, \\
    \dot Q(t) &= \Gamma(t)' Q_{c} \Gamma(t), \qquad && Q(0) = 0,  \\
    \dot M(t) &= -\Gamma(t)' Q_{c}, \qquad && M(0) = 0,
\end{alignat}
where
\begin{equation}
    \Gamma(t) = \begin{bmatrix} C_c & D_c \end{bmatrix} \begin{bmatrix} A(t) & B(t) \\ 0 & I \end{bmatrix},
\label{eq:gamma(t)}
\end{equation}
\end{subequations}
may be used to compute ($A = A(T_s)$, $B=B(T_s)$, $Q=Q(T_s)$, $ M=M(T_s)$). 
% may be used to compute $(A,B,Q,M)$ as
% \begin{align}
%     A = A(T_s), \quad 
%     B = B(T_s), \quad 
%     Q = Q(T_s), \quad 
%     M = M(T_s).
% \end{align}
{\flushright \hfill $\blacksquare$}
\end{proposition}
%% %%%%%%%%%%%%%%%%%%%%%%%%%%
% \begin{remark} \label{remark:AlternativeBt}
% An alternative expression of $B(t)$ is 
% \begin{equation}
%     \dot B(t) = A(t) B_c = A_cB(t) + B_c, \qquad B(0) = 0.
% \end{equation}
% In this case, we will use the first expression since it is easier to solve than the latter. The proof can be found in Appendix~\ref{app:DistributionProofs}.
% % Note that the case $C_c = \begin{bmatrix} I \\ 0 \end{bmatrix}$, $D_{c} = \begin{bmatrix} 0  \\ I \end{bmatrix}$, $Q_{c} = \begin{bmatrix} Q_{c,xx} & 0 \\ 0 & Q_{c,uu}\end{bmatrix}$, and $\bar{z}(t) = [\bar{x}(t); \, \bar{u}(t)]$ corresponds to
% % \begin{equation}
% % \begin{split}
% %     l_c(\tilde z(t)) &= \frac{1}{2} \left[x(t) - \bar x(t)\right]' Q_{c,xx} \left[ x(t) - \bar x(t) \right] 
% %     \\ & \qquad + \frac{1}{2} \left[ u(t) - \bar u(t) \right]' Q_{c,uu} \left[u(t)-\bar u(t) \right].
% % \end{split}
% % \end{equation}
% {\flushright \hfill $\blacksquare$}
% \end{remark}
%% %%%%%%%%%%%%%%%%%%
\subsection{Certainty equivalent LQ control for a stochastic system}
Consider an initial state and an input noise modelled by the following random variables,
\begin{align}
    {\bs x}(t_0) & \sim N(\hat x_0,P_0), \qquad 
    d{\bs \omega} (t) \sim N_{iid}(0, I dt).
\end{align}
The stochastic system can be described as continuous-time linear stochastic differential equations (SDEs) in the form
\begin{subequations}
\begin{alignat}{3}\label{eq:linearSDE}
    d {\bs x}(t) &= \left( A_c {\bs x}(t) + B_c u(t) \right) dt + G_c d{\bs \omega}(t), \\
    {\bs z}(t) &= C_c {\bs x}(t) + D_c  u(t).
\end{alignat}
% with piecewise constant inputs 
% \begin{equation}
%     u(t) = u_k \qquad t_k \leq t < t_{k+1}
% \end{equation}
\end{subequations}
The corresponding discrete-time stochastic system is 
\begin{subequations}
\begin{alignat}{3}
    {\bs x}_{k+1} &= A {\bs x_k} + B u_k + {\bs w}_k, \\
    {\bs z}_k &= C {\bs x_k} +  D u_k,
\end{alignat}
where
\begin{align}
    {\bs x}_0 &\sim N(\hat{x}_0,P_0), \qquad {\bs w}_k \sim N_{iid}(0,R_{ww}).
\end{align}
\end{subequations}

\begin{proposition}[Discretization of the linear SDE]
\label{prop:DiscretizationoftheLinearSDE}
The system of differential equations
\begin{subequations}
\begin{alignat}{3}
    \dot A(t) &= A_c A(t), \qquad && A(0) = I, \\
    \dot B(t) &= A(t) B_c, \qquad && B(0) = 0, \\
    \dot R_{ww} &= \Phi(t) \Phi(t)', \qquad && R_{ww}(0) = 0,
\end{alignat}
where 
\begin{equation}
    \Phi(t) = A(t) G_c,
\end{equation}
\end{subequations}
can be used to compute ($A=A(T_s)$, $B=B(T_s)$, $R_{ww}=R_{ww}(T_s)$).
% may be used to compute ($A$,$B$,$R_{ww}$,$C$,$D$) as
% \begin{subequations}
% \begin{align}
%     A = A(T_s), & \qquad 
%     B = B(T_s), \qquad 
%     R_{ww} = R_{ww}(T_s), \\
%     & C = C_c, \qquad  \qquad  D = D_c.
% \end{align}
% \end{subequations}
{\flushright \hfill $\blacksquare$}
\end{proposition}

\subsection{Stochastic linear-quadratic optimal control problem}
Consider the stochastic LQ-OCP
\begin{subequations}
\begin{alignat}{5}
\label{eq:Cont-Stc-LQ-OCP}
& \min_{{\bs x},u,{\bs z},{\tilde{\bs z}}}  \quad && \psi = E \set{ \phi = \int_{t_0}^{t_0+T} l_c({\tilde{\bs z}}(t))dt } \\
& s.t. && {\bs x}(t_0) \sim N(\hat x_0, P_0), \\
& && d{\bs \omega}(t) \sim N_{iid}(0,I dt), \\
& && u(t) = u_k, \quad t_k \leq t < t_{k+1}, \, k \in \mathcal{N}, \\
& && d {\bs x}(t) = (A_c {\bs x}(t) + B_c u(t))dt + G_c d{\bs \omega}(t), \\
& && {\bs z}(t) = C_c {\bs x}(t) + D_c u(t), \\
& && {\bar z}(t) = \bar z_k, \quad t_k \leq t < t_{k+1}, \, k \in \mathcal{N}, \\
& && {\tilde {\bs z}}(t) = {\bs z}(t) - \bar{z}(t).
\end{alignat}
\end{subequations}
The corresponding discrete-time stochastic LQ-OCP is 
\begin{subequations}
\begin{alignat}{5}
\label{eq:Dist-Stc-LQ-OCP}
& \min_{ x,u} \quad  && \psi = E\set{\phi = \sum_{k \in \mathcal{N}} l_k({\bs x}_k,u_k)+  l_{s,k} (\bs{x}_k,u_k)}  \\ 
& s.t. && {\bs x}_0 \sim N(\hat x_0, P_0), \\
& && {\bs w_k} \sim N_{iid}(0,R_{ww}), \qquad \qquad \: k \in \mathcal{N}, \\
& && {\bs x}_{k+1} = A {\bs x}_k + B u_k + {\bs w}_k, \qquad k \in \mathcal{N}, 
\end{alignat}
\end{subequations}
where the stage cost function $l_k(\bs{x}_k, u_k)$ is
\begin{subequations}
    \begin{equation}
    l_k(\bs{x}_k,u_k) = \frac{1}{2} \begin{bmatrix} \bs{x}_k \\ u_k \end{bmatrix}' Q \begin{bmatrix} \bs{x}_k \\ u_k \end{bmatrix} + q_k' \begin{bmatrix} \bs{x}_k \\ u_k \end{bmatrix} + \rho_k,
    \label{eq:deterministicStageCostofStochasticCosts}
    \end{equation}
and the stochastic stage cost function $l_{s,k}(\bs{x}_k, u_k)$ is
    \begin{equation}
        l_{s,k}(\bs{x}_k, u_k) = \int_{t_k}^{t_{k+1}} \frac{1}{2} \bs{w}(t)' Q_{c,ww} \bs{w}(t) + \bs{q}_{s,k}' \bs{w}(t) dt.
    \end{equation}
\end{subequations}
$Q$, $q_k$, and $\rho_k$ in \eqref{eq:deterministicStageCostofStochasticCosts} are identical to the deterministic case. The state variables and system matrices of $l_{s,k}(\bs{x}_k,u_k)$ are
\begin{subequations}
    \begin{align}
        & \bs{w}(t) = \int_{0}^{t} A(s) G_c d\bs{\omega}(s),  \quad 
        && Q_{c,ww} = C_c' Q_{c,xx} C_c,  
        \\
        & \Tilde{\bs{z}}_k = \Gamma(t) \begin{bmatrix} \bs{x}_k \\ u_k \end{bmatrix} - \bar{z}_k, 
        && \bs{q}_{s,k} = \begin{bmatrix}
            C_c' & 0
        \end{bmatrix} Q_{c} \Tilde{\bs{z}}_k.
    \end{align}
\end{subequations}
The expectation of the stochastic LQ-OCP is~\cite{Karl1970IntroToStoContTheory} 
\begin{subequations}
\begin{alignat}{5}
& \min_{ x,u } \: && \psi = \sum_{k \in \mathcal{N}} l_k(x_k,u_k) + \frac{1}{2} \left[ \text{tr}\left( Q \bar{P}_k \right) + \int_{t_k}^{t_{k+1}} \text{tr} \left(Q_{c,ww} P_{w} \right) dt \right]\\
& s.t. && x_0 = \hat x_0, \\
& &&  x_{k+1} = A x_k + B u_k, \quad k \in \mathcal{N}, 
\end{alignat}
\label{eq:analyticMeanOfStochasticCosts}
\end{subequations}
where 
\begin{subequations}
    \begin{equation}
    \begin{bmatrix}
            \bs{x}_k \\ u_k
        \end{bmatrix} \sim N(m_k, \bar{P}_k),
    \quad 
    m_k = \begin{bmatrix}
        x_k \\ u_k
    \end{bmatrix},
    \quad 
    \bar{P}_k = \begin{bmatrix} P_k & 0 \\ 0 & 0 \end{bmatrix},
\end{equation}
\begin{equation}
    P_{k+1} = A P_k A' + R_{ww}, \qquad P_w = \text{Cov} \left(\bs{w}(t)\right).
\end{equation}
\end{subequations}

% The mean of the stochastic linear quadratic optimal control problem is 
% \begin{subequations}
% \begin{alignat}{5}
% & \min_{x,u} \quad && \psi = \sum_{k \in \mathcal{N}} l_k(x_k,u_k) + \psi_{tr} \\
% & s.t. && x_0 = \hat x_0 \\
% & &&  x_{k+1} = A x_k + B u_k \quad && k \in \mathcal{N} 
% \end{alignat}
% \end{subequations}
% where 
% \begin{equation}
%     \psi_{tr} = \text{tr}(Q_c (C_c P_0 C_c')) + \int_{t_0}^{t_0+T} \text{tr} \left( Q_c R_{zz}(t) \right) dt
% \end{equation}
\begin{proposition}[Discretization of the stochastic LQ-OCP]
\label{prop:DiscretizationoftheStochasticLQ-OCP}
The system of differential equations
\begin{subequations}
\label{eq:StochasticLQG:MatrixODEsystem}
\begin{alignat}{3}
\dot A(t) &= A_c A(t), \qquad && A(0) = I, \\
\dot B(t) &= A(t) B_c, \qquad && B(0) = 0 ,\\
\dot R_{ww}(t) &= \Phi(t) \Phi(t)', \qquad && R_{ww}(0) = 0, \\
\dot Q(t) &= \Gamma(t)' Q_{c} \Gamma(t) , && Q(0) = 0 ,\\
\dot M(t) &= -\Gamma(t)' Q_{c}, \qquad && M(0) = 0, 
% \\ \dot Q_{ww}(t) &= \Lambda(t)' Q_{c,zz} \Lambda(t), \qquad && Q_{ww}(0) = 0 ,
\end{alignat}
where
\begin{align}
    \Phi(t) &= A(t) G_c,\qquad \Gamma(t) = \begin{bmatrix} C_c & D_c \end{bmatrix} \begin{bmatrix} A(t) & B(t) \\ 0 & I \end{bmatrix},  
    % \Lambda(t) &= C_c \Phi(t) 
 % R_{zz}(t) &= \Lambda(t) \Lambda(t)' 
\end{align}
\end{subequations}
can be used to compute ($A=A(T_s)$, $B=B(T_s)$, $R_{ww}=R_{ww}(T_s)$, $Q=Q(T_s)$, $M=M(T_s)$).
%\begin{subequations}
% \begin{align}
%     A = A(T_s),   
%      B = B(T_s), 
%     R_{ww} = R_{ww}(T_s), 
%     Q = Q(T_s), 
%     M = M(T_s). 
% \end{align}
% \label{eq:systemOfDifferentialEquations}
%\end{subequations}
{\flushright \hfill $\blacksquare$}
\end{proposition}

\begin{proposition}[Distribution of the stochastic costs] \label{prop:DistributionofStochasticCost}
Using Euler-Maruyama (EM) discretization, we can reformulate~\eqref{eq:Dist-Stc-LQ-OCP} into isolated stochastic form as
\begin{subequations}
    \begin{equation}
        \begin{split}
            \phi &= \sum_{k \in \mathcal{N}} l_k( \bs{x}_k, u_k) +  l_{s,k}(\bs{x}_k, u_k)
            \\
            &= \frac{1}{2} \begin{bmatrix} \bs{x}_0 \\ \bs{W}_{\mathcal{N}} \end{bmatrix}' Q_{\mathcal{N}} \begin{bmatrix} \bs{x}_0 \\ \bs{W}_{\mathcal{N}} \end{bmatrix} + q_{\mathcal{N}}' \begin{bmatrix} \bs{x}_0 \\ \bs{W}_{\mathcal{N}} \end{bmatrix} + \rho_{\mathcal{N}},
        \end{split}
    \label{eq:EMReformulatedQP}
    \end{equation}
where 
    \begin{equation}
    \begin{bmatrix}
      \bs{x}_0 \\ \bs{W}_{\mathcal{N}}
    \end{bmatrix}
    \sim N(\bar{m}, \bar{P}), \quad \bar{m} = \begin{bmatrix}
        x_0 \\ 0
    \end{bmatrix},
    \quad 
    \bar{P} = \begin{bmatrix}
    P_0 & 0 \\ 0 & P_w
  \end{bmatrix}.
\end{equation}
\end{subequations}
$\bs{W}_{\mathcal{N}}$ is a vector of sub-sampling random variables over the horizon, and its covariance is $P_w = \text{diag}([I\delta t, I\delta t, \ldots, I \delta t])$.

Based on \cite{Das2021genchi2}, the reformulated stochastic cost follows a generalized $\chi^2$ distribution, and its expectation and variance can be expressed as 
\begin{subequations}\label{eq:meanAndVariance}
    \begin{alignat}{3}
        & E\set{\phi} = \frac{1}{2}\bar{m}'Q_{\mathcal{N}} \bar m + q_{\mathcal{N}}'\bar m + \rho_{\mathcal{N}} + \frac{1}{2}\text{tr}\left( Q_{\mathcal{N}} \bar P\right),
        \label{eq:meanOfStochasticCosts} 
        \\ 
        \begin{split}
        & V\set{\phi} = q_{\mathcal{N}}' \bar P q_{\mathcal{N}} + 2\bar m 'Q_{\mathcal{N}} \bar P q_{\mathcal{N}} + \bar{m}'Q_{\mathcal{N}} \bar P Q_{\mathcal{N}} \bar m \\
        & \qquad \qquad + \frac{1}{2}\text{tr}(Q_{\mathcal{N}} \bar P Q_{\mathcal{N}} \bar P).
        \label{eq:varOfStochasticCosts}
        \end{split}
    \end{alignat}
\end{subequations}
The stochastic cost $\phi$ in \eqref{eq:Cont-Stc-LQ-OCP} follows a generalized $\chi^2$ distribution with mean and variance given by \eqref{eq:meanAndVariance} when taking the limit of integration steps $n \rightarrow \infty$. See Appendix \ref{app:DistributionProofs} for more details.

{\flushright \hfill $\blacksquare$}
\end{proposition}
%  has the same format as \eqref{eq:deterministic-stageCost}, but with the stochastic system state $\bs{x}_k$.
 
%     \begin{equation}
%         q_k = M \bar{z}_k
%     \end{equation}
    
%     \begin{equation}
%         \rho_k = \int_{t_k}^{t_{k+1}} l_c(\bar z_k) dt = l_c(\bar z_k) T_s
%     \end{equation}
% and the stage cost function $l_{s,k}(\bs{x}_k, u_k)$
%     \begin{equation}
%         l_{s,k}(\bs{x}_k, u_k) = \int_{t_k}^{t_{k+1}} \frac{1}{2} \bs{W}(t)' Q_{w} \bs{W}(t) + \bs{q}_{s,k}' \bs{W}(t) dt
%     \end{equation}

%     \begin{equation}
%         \bs{W}(t) = \int_{0}^{t} A(s)G_c d\omega(s) 
%     \end{equation}

%% file: tex/NumericalMethodsZHAZ.tex
\section{Numerical Methods of LQ Discretization}
\label{sec:NumericalMethods}
This section introduces numerical methods for the discretization of continuous-time LQ-OCPs.
\subsection{Ordinary differential equation methods}
\begin{table}[b]
    \centering
    \caption{$\Lambda$ and $\Theta$ of the ODE method with discretization methods}%
    \begin{tabular}{ p{1.45cm} p{2.94cm}  p{3.075cm}  }
    \hline
    {Methods} & \makecell{$\Lambda$}  & \makecell{$\Theta$} \\ 
    \hline
    {Expl. Euler}  &  \makecell{$I+hA_c$}         & \makecell{$I$}   \\
    {Impl. Euler}  &  \makecell{$(I-hA_c)^{-1}$}  & \makecell{$(I-hA_c)^{-1}$}\\
    {Expl. Trape.} &  \makecell{$I+hA_c+0.5h^2A_c^2$} & \makecell{$I+0.5hA_c$} \\
    {Impl. Trape.} &  \makecell{$(I+0.5hA_c)^{-1}$ \\ $(I-0.5hA_c)$} & \makecell{$(I-0.5hA_c)^{-1}$}\\
    {ESDIRK34}  &  \makecell{$(I-0.44hA_c)^{-3}$ \\ $(I-0.31hA_c-$ $0.24h^2A_c^2)$}  & \makecell{$(I-0.44hAc)^{-3}$ \\ $(I-0.81hAc+$ $0.08h^2Ac^2)$}\\
    {Classic RK4} &  \makecell{$I+hA_c+0.5h^2A_c^2+$ \\ $0.17h^3A_c^3+$ $0.04h^4A_c^4$} & \makecell{$I+0.5hA_c+0.17h^2A_c^2+$ \\ $0.04h^3A_c^3$}  \\
    \hline
    \end{tabular}
    \label{tab:differentDiscretizationMethods}%
\end{table}%
Consider an s-stage ODE method with $N \in \mathbb{Z}^+$ integration steps and the time step $h = \frac{T_s}{N}$. We compute ($A$, $B$, $R_{ww}$, $Q$, $M$) as 
\begin{subequations}
    \label{eq:ODEmethods-numericalExpressions}
    \begin{align}
         A_{k+1} &= \Lambda A_k,     \: && k \in \mathcal{N},\\
         B_{k+1} &= B_k + \Theta A_k \bar B_c,  \: && k \in \mathcal{N}, \\
         \Gamma_{k+1} &= \Omega \Gamma_k, \: && k \in \mathcal{N},
         \\
         M_{k+1} &= M_k + \Gamma_k' \bar{M}_c,\: && k \in \mathcal{N},
         \\
         Q_{k+1} &= Q_k + \Gamma_k' \bar{Q}_{c} \Gamma_k, \: && k \in \mathcal{N},
         \\
         R_{ww,k+1} &=  R_{ww,k} + \sum_{i=1}^{s} b_i \Lambda_i A_{k} \bar{R}_{ww,c} A_{k}' \Lambda_i',
          \: && k \in \mathcal{N},
    \end{align}
\end{subequations} 
where $\bar B_c = h B_c$, $H=\begin{bmatrix} C_c & D_c \end{bmatrix}$, $\bar{M}_c= - h\sum_{i=1}^{s} b_i \Omega_i' H' Q_c$, $\bar{Q}_c = h\sum_{i=1}^{s} b_i \Omega_i' H' Q_c H\Omega_i$, and $\bar{R}_{ww,c} = h G_c G_c'$ are constant matrices. $\Lambda$, $\Theta$, $\Omega$ are functions of the coefficients, $a_{i,j}$ and $b_i$ for $i=1,2,\ldots,s$ and $j=1,2,\ldots,s$ in the Butcher tableau of the ODE method. They are computed as 
\begin{subequations}
    \label{eq:ODEmethods-ConstantCoefficients}
    \begin{alignat}{5}
        A_{k,i} &= A_k + h\sum_{j=1}^{s} a_{i, j} \dot A_{k,j} = \Lambda_{i} A_k,
        \\
        B_{k,i} &= B_k + h\sum_{j=1}^{s} a_{i,j} \dot B_{k,j} = B_k + \Theta_{i} A_k \bar B_c,
        \\
        \Gamma_{k, i} &= \Omega_i \Gamma_k = \begin{bmatrix}
                \Lambda_i & \Theta_i  \bar{B}_c\\ 
                0 & I \end{bmatrix} \Gamma_k, 
        \\
        \Lambda &= I + h \sum_{i=1}^s b_i A_c \Lambda_i,  \\
        \Theta &= \sum_{i=1}^s b_i \Lambda_{i},       
        \\
        \Omega &= \begin{bmatrix}
                \Lambda & \Theta \bar{B}_c \\ 
                0 & I \end{bmatrix},
    \end{alignat}
\end{subequations}
where $\Lambda_i$, $\Theta_i$, $\Omega_i$ are coefficients of stage variables $A_{k,i}$, $B_{k,i}$, $\Gamma_{k, i}$. 

Consequently, the differential equation systems $A(T_s)=A_N$, $B(T_s)=B_N$, $R_{ww}(T_s)=R_{ww, N}$, $Q(T_s)=Q_N$, $M(T_s)=M_N$, has constant coefficients $\Lambda_{i}$, $\Theta_{i}$, $\Omega_i$, $\Lambda$, $\Theta$ and $\Omega$ when using 
fixed-time-step ODE methods. The constant coefficients can be computed offline. Table \ref{tab:differentDiscretizationMethods} describes $\Lambda$ and $\Theta$ for different discretization methods. Algorithm \ref{alg:ODEmethod-LQDiscretization} presents the ODE methods for LQ discretization.

% Consequently, we can solve the differential equation systems $A(T_s)=A_N$, $B(T_s)=B_N$, $R_{ww}(T_s)=R_{ww, N}$, $Q(T_s)=Q_N$, $M(T_s)=M_N$. The coefficients $\Lambda_{i}$, $\Theta_{i}$, $\Omega_i$, $\Lambda$, $\Theta$ and $\Omega$ are constant when using fixed-time-step ODE methods, thus we can compute them offline. Table \ref{tab:differentDiscretizationMethods} describes $\Lambda$ and $\Theta$ for different discretization methods. Algorithm \ref{alg:ODEmethod-LQDiscretization} presents the ODE methods for LQ discretization.
% \vspace{}
\begin{algorithm}[tb]
\caption{ODE method for LQ Discretization}
\label{alg:ODEmethod-LQDiscretization}
\begin{flushleft}
    \textbf{Input:} $(A_c, B_c, G_c, C_c, D_c, Q_c, T_s, h)$ \\
    \textbf{Output:} $(A(T_s),B(T_s),C,D,Q(T_s),M(T_s),R_{ww}(T_s))$ 
\end{flushleft}
\begin{algorithmic}
\State Set initial states \\
        ($k=0$, $A_k = I$, $B_k = 0$, $Q_k = 0$, $M_k = 0$, $R_{ww, k} = 0$)
\State Use \eqref{eq:ODEmethods-ConstantCoefficients} to compute  ($\Lambda_i$, $\Theta_i$, $\Omega_i$, $\Lambda$, $\Theta$, $\Omega$)
\State Compute integration steps $N = \frac{T_s}{h}$
\While{$k < N$}
    \State {Use \eqref{eq:ODEmethods-numericalExpressions} to update $(A_{k}, B_{k}, \Gamma_{k}, Q_{k}, M_{k}, R_{ww, k})$} 
    \State {Set $k \leftarrow k + 1$}
\EndWhile

% \While{$t < T_s$} 
% \If{$t+h > T_s$}
%     \State Update $h = T_s - t$
%     \State Use \eqref{eq:ODEmethods-ConstantCoefficients02} to compute  ($\Lambda_i$, $\Theta_i$, $\Omega_i$)
%     \State Use \eqref{eq:ODEmethods-ConstantCoefficients01} to compute 
%     ($\Lambda$, $\Theta$, $\Omega$, $\bar{B}_c$, $\bar{M}_c$, $\bar{Q}_c$, $\bar{R}_{ww,c}$) 
% \EndIf
% \State {Use \eqref{eq:ODEmethods-numericalExpressions} to update $(A_{k}, B_{k}, \Gamma_{k}, Q_{k}, M_{k}, R_{ww, k})$} 
% \State Set $t = t + h$
% \EndWhile
\State Get system matrices
 $(A(T_s) = A_k, B(T_s)= B_k, C = C_c, D = D_c, Q(T_s) = Q_k, M(T_s) = M_k, R_{ww}(T_s) = R_{ww, k})$
\end{algorithmic}
\end{algorithm}

%% %%%%%%%%%%%%%%%%%%%%%%%%%%%%%%%%%%%%%%%%%%%%%%%%
\subsection{Matrix exponential method}
The matrix exponential method describes the LQ discretization by three matrix exponential problems
\begin{subequations}
    \begin{equation}
        \begin{split}
            \begin{bmatrix}
                \Phi_{1, 11} & \Phi_{1, 12}\\
                0 & \Phi_{1,22}
            \end{bmatrix} &= \text{exp} \left(\begin{bmatrix}
                                -H' & \bar Q_c \\
                                0   & H
                            \end{bmatrix}t \right), \\
        \end{split}
    \end{equation}
    \begin{equation}
        \begin{split}
            \begin{bmatrix}
                I & \Phi_{2, 12}\\
                0 & \Phi_{2,22}
            \end{bmatrix} &= \text{exp} \left(\begin{bmatrix}
                                0 & I \\
                                0 & H'
                            \end{bmatrix}t \right), \\
        \end{split}
    \end{equation}
    \begin{equation}
        \begin{split}
            \begin{bmatrix}
                \Phi_{3, 11} & \Phi_{3, 12}\\
                    0 & \Phi_{3,22}
            \end{bmatrix} &= \text{exp} \left(\begin{bmatrix}
                                -A_c & \bar{G}_c \\
                                0   & A_c'
                            \end{bmatrix} t \right), \\
        \end{split}
    \end{equation}
\end{subequations}
where 
\begin{subequations}
    \begin{align}
        H &= \begin{bmatrix} A_c & B_c \\ 0 & 0 \end{bmatrix}, \qquad 
        && \bar{M}_c = -\begin{bmatrix}  C_c & D_c \end{bmatrix}' Q_c,
        \\
        \bar{Q}_c &= - \bar{M}_c \begin{bmatrix} C_c & D_c \end{bmatrix}, \qquad 
        && \bar{G}_c = G_c G_c'.
    \end{align}
\end{subequations}
The elements of matrix exponential problems are 
\begin{subequations}
    \begin{align}
        \begin{split}
            \Phi_{1,22} &= \Gamma(t) = \begin{bmatrix}
            A(t) & B(t) \\ 0 & I
            \end{bmatrix} ,
        \end{split}\\
        \begin{split}
            \Phi_{1,12} & = \Gamma(-t)' \int_{0}^{t} \Gamma(\tau)' \bar{Q}_c \Gamma(\tau) d\tau,
        \end{split}\\
        \begin{split}
            \Phi_{2,12} &= \int_{0}^{t} \Gamma(\tau)' d\tau,
        \end{split}\\
        \begin{split}
            \Phi_{3,22} & = A(t)',
        \end{split}\\
        \begin{split}
            \Phi_{3,12} &= A(-t)' \int^{t}_{0} A(\tau) \bar{G}_c A(\tau)' d \tau.
        \end{split}
    \end{align}
\end{subequations}
Set $t = T_s$, we can compute differential equations $A$, $B$, $R_{ww}$, $Q$ and $M$ as
\begin{subequations}
    \begin{align}
        A(T_s) &= \Phi_{1,22}(1:n_x, 1:n_x), \\
        B(T_s) &= \Phi_{1,22}(1:n_x, n_x+1:\text{end}), \\
        Q(T_s) &= \Phi_{1,22}'\Phi_{1,12}, \\
        M(T_s) &= \Phi_{2,12} \bar{M}_c, \\
        R_{ww}(T_s) &= \Phi_{3,22}' \Phi_{3,12}.
    \end{align}
\end{subequations}
The matrix exponential method is inspired by formulas from~\cite{Moler1978NineteenDubiousWays,Moler2003NineteenDubiousWays25YearsLater,vanLoan1978MatrixExponential}.
% 

%% Step doubling method %%%%%%%%%%%%%%%%%%%%%%%
\subsection{Step-doubling method}
Consider an s-stage ODE method with $N \in \mathbb{Z}^+$ integration steps and the time step $h = \frac{T_s}{N}$. The matrices
\begin{subequations}
    \begin{align}
        \Tilde{A}(N) &= \Lambda^N,                       && \Tilde{A}(1) = \Lambda, \\
        \Tilde{B}(N) &= \displaystyle \sum_{i=0}^{N-1} \Lambda^i,  && \Tilde{B}(1) = I,       \\
        \Tilde{\Gamma}(N) &= \Omega^N ,                 && \Tilde{\Gamma}(1) = I_{xu},    \\ 
        \Tilde{M}(N) &= \displaystyle \sum_{i=0}^{N-1} \left(\Gamma^i \right)' , && \Tilde{M}(1) = I_{xu}, \\
        \Tilde{Q}(N) &= \displaystyle \sum_{i=0}^{N-1} \left(\Gamma^i \right)' \bar{Q}_c \left(\Gamma^i \right), && \Tilde{Q}(1) = \bar{Q}_c, \\
        \Tilde{R}(N) &= \displaystyle \sum_{i=0}^{N-1} \left(A^i \right) \bar{R}_{ww,c} \left(A^i \right)', && \Tilde{R}(1) = \bar{R}_{ww,c},
    \end{align}
    \label{eq:Stepdoubling_finalexpressions}
\end{subequations}
can be used to solve $(A,B,M,Q,R_{ww})$
\begin{subequations}
    \begin{align}
        A(T_s) &= \Tilde{A}(N), \\
        B(T_s) &= \Theta \Tilde{B}(N) \bar{B}_c, \\ 
        M(T_s) &= \Tilde{M}(N) \bar{M}_c ,\\ 
        Q(T_s) &= \Tilde{Q}(N),  \\
        R_{ww}(T_s) &=  \sum_{i=1}^s b_i \Lambda_i \Tilde{R}(N) \Lambda_i',
    \end{align}
    \label{eq:Stepdoubling_numericalExpressions}%
\end{subequations}
where $\Lambda$, $\Theta$, $\Omega$, $\bar{B}_c$, $\bar{M}_c$ and $\bar{Q}_c$ are the same as in the ODE method case. $I_{xu} \in \mathbb{R}^{n_{xu} \times n_{xu}}$ is an identity matrix with the size $n_{xu} = n_x + n_u$. 
\begin{figure}[bt]
    \centering
    \includegraphics[width=0.5\textwidth]{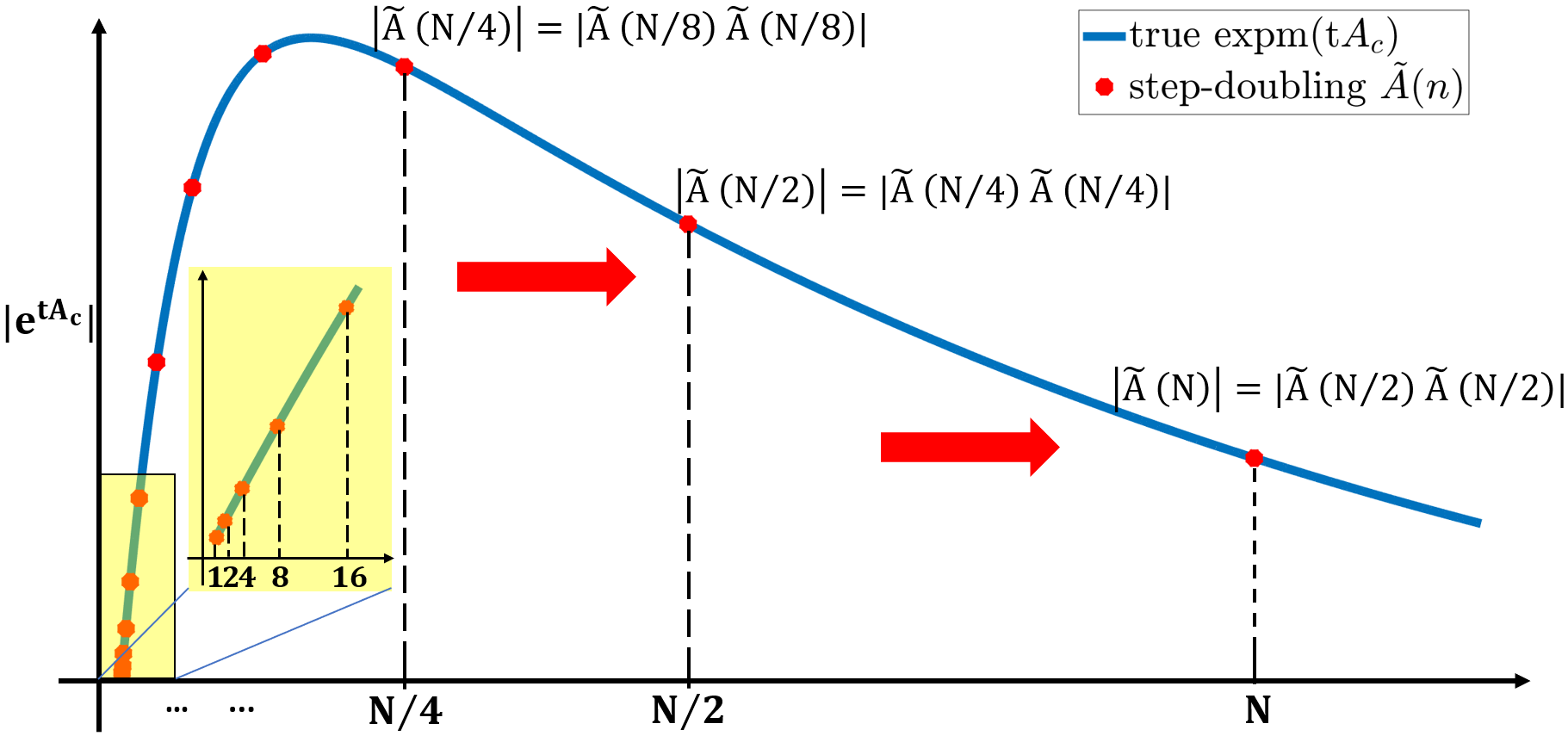}
    \caption{The exponential of a matrix $A_c$. The blue line is the true result computed with expm() in MATLAB. The red dots are the results of the step-doubling method.}
    \label{fig:Stepdoubling_Plots}
\end{figure}

Fig. \ref{fig:Stepdoubling_Plots} shows an example of $\Tilde{A}(t)=e^{t A_c}$, where the step-doubling method (red dots) uses the $\frac{n}{2}^{th}$ step result as the initial state to compute the double step's result $\Tilde{A}(n)$. We then get the step-doubling expression as 
\begin{subequations}
    \begin{equation}
    \Tilde{A}(1) \rightarrow \Tilde{A}(2) \rightarrow \Tilde{A}(4) \rightarrow \ldots \rightarrow \Tilde{A}(\frac{N}{4})\rightarrow \Tilde{A}(\frac{N}{2}) \rightarrow \Tilde{A}(N),
    \label{eq:Stepdoubling_exampleofA(t)}
    \end{equation}
    where
    \begin{equation}
        \Tilde{A}(n) = \Tilde{A}(\frac{n}{2}) \Tilde{A}(\frac{n}{2}), \qquad \qquad n \in [2,4, \ldots, \frac{N}{2}, N].
    \end{equation}
    \label{eq:Stepdoubling_numericalExpressionOfA(t)}%
\end{subequations}%
Eq.~\eqref{eq:Stepdoubling_numericalExpressionOfA(t)} is inspired by the scaling and squaring algorithm for solving matrix exponential problem presented in~\cite{NewStepDoubling2010Nigham,ExponentialIntegrators2011Higham,ScalingSquaringRevisited2005Nigham}. We apply the same idea for other differential equations. Table \ref{tab:Stepdoubling-numericalExpressions} describes the step-doubling expressions for ($\Tilde{A}$, $\Tilde{B}$, $\Tilde{\Gamma}$, $\Tilde{Q}$, $\Tilde{M}$, $\Tilde{R}_{ww}$). The step-doubling method takes only $j$ steps to get the same result as the ODE method with $N=2^j$ integration steps. Algorithm \ref{algo:Stepdoubling-LQDiscretization} describes the step-doubling method for LQ discretization.
\begin{table}[tb]
    \centering
    \caption{Numerical expressions of the step-doubling method}%
    \label{tab:Stepdoubling-numericalExpressions}%
    \begin{tabular}{ p{1.1cm} p{2.6cm}  p{3.0cm}  }
    \hline
    {ODEs}  & {ODE expression}  & {Step-doubling expression}    
    \\ \hline
    {$\Tilde{A}(N)$} & {$\Lambda^N$}  & {$\Tilde{A}(\frac{N}{2}) \Tilde{A}(\frac{N}{2})$} 
    \\
    {$\Tilde{B}(N)$} & {$\displaystyle \sum_{i=0}^{N-1} \Lambda^i$}  & {$\Tilde{B}(\frac{N}{2}) \left(I + \Tilde{A}(\frac{N}{2}) \right)$} 
    \\
    {$\Tilde{\Gamma}(N)$} & {$\Omega^N$} & {$\Tilde{\Gamma}(\frac{N}{2}) \Tilde{\Gamma}(\frac{N}{2})$} 
    \\
    {$\Tilde{M}(N)$} & {$\displaystyle \sum_{i=0}^{N-1} \Omega^i \bar{M}_c$} & {$\Tilde{M}(\frac{N}{2}) \left(I + \Tilde{\Gamma}(\frac{N}{2})' \right)$}
    \\
    {$\Tilde{Q}(N)$} & {$\displaystyle\sum_{i=0}^{N-1} \left( \Gamma^i  \right)' \bar{Q}_c \left( \Gamma^i \right)$} & {$\Tilde{Q}(\frac{N}{2}) + \Tilde{\Gamma}(\frac{N}{2})' \Tilde{Q}(\frac{N}{2}) \Tilde{\Gamma}(\frac{N}{2})$} 
    \\
    {$\Tilde{R}(N)$} & {$\displaystyle \sum_{i=0}^{N-1} \left( A^i  \right) \bar{R}_{ww,c} \left( A^i \right)'$}           & {$\Tilde{R}(\frac{N}{2}) + \Tilde{A}(\frac{N}{2}) \Tilde{R}(\frac{N}{2}) \Tilde{A}(\frac{N}{2})'$} \\ \hline
    \end{tabular}%
\end{table}%
\begin{algorithm}[tb]
\caption{Step-doubling method for LQ Discretization}
\label{algo:Stepdoubling-LQDiscretization}
\begin{flushleft}
    \textbf{Input:} $(A_c, B_c, G_c, C_c, D_c, Q_c, T_s, j)$ \\
    \textbf{Output:} $(A(T_s),B(T_s),C,D,Q(T_s),M(T_s),R_{ww}(T_s))$ 
\end{flushleft}
\begin{algorithmic}
\State Compute the number of step $N = 2^j$
\State Compute the step size $h = \frac{T_s}{N}$
\State Use \eqref{eq:ODEmethods-ConstantCoefficients} to compute  ($\Lambda_i$, $\Theta_i$, $\Omega_i$, $\Lambda$, $\Theta$, $\Omega$)
\State Set initial states of step doubling matrices ($i=1, \Tilde{A}(i) = \Lambda$, $\Tilde{B}(i) = I$, $\Tilde{Q}(i) = \bar{Q}_c$, $\Tilde{M}(i) = I_{xu}$, $\Tilde{R}(i) = \bar{R}_{ww,c}$)
\While{$i \leq j$} 
    \State Use equations from Table \ref{tab:Stepdoubling-numericalExpressions} to update ($\Tilde{\Gamma}(i)$,$\Tilde{M}(i)$, $\Tilde{Q}(i)$,$\Tilde{R}(i)$)
    \State Use equations from Table \ref{tab:Stepdoubling-numericalExpressions} to update ($\Tilde{A}(i)$, $\Tilde{B}(i)$) 
    \State Set $i = i + 1$
\EndWhile
\State Use \eqref{eq:Stepdoubling_numericalExpressions} to compute 
 $(A(T_s), B(T_s), Q(T_s), M(T_s), R_{ww}(T_s))$
\end{algorithmic}
\end{algorithm}

%% file: tex/NumericalExperiments.tex
\section{Numerical Experiments}
\label{sec:NumericalExperiments}
% In this section, we test and compare the proposed numerical methods and investigate the distribution of the stochastic costs via Monte Carlo simulations. 
Consider a continuous-time stochastic LQ-OCP with system matrices
\begin{equation}
    A_c = \begin{bmatrix} -49 & 24 \\ -64 & 31 \end{bmatrix}, \quad  B_c = \begin{bmatrix} 2 & 0.5 \\ 1 & 3 \end{bmatrix}, \quad G_c = \begin{bmatrix} 0.1 & 0.0 \\ 0.0 & 0.1 \end{bmatrix}.
\end{equation}
The system output matrices are $C_c=[1.0, 1.0]$, $D_c=[0.0,0.0]$, and the system references and system inputs are
\begin{subequations}
    \begin{align}
        &\bar{z}(t) = \bar{z}_k = 3.0,  && t_k \leq t \leq t_{k+1}, \quad k\in \mathcal{N}, \\
        &\bar{u}(t) = \bar{u}_k = 0.0,  && t_k \leq t \leq t_{k+1}, \quad k\in \mathcal{N}, \\
        &u(t)= u_k = \begin{bmatrix} 1.0 & 1.0 \end{bmatrix}^T ,&& t_k \leq t \leq t_{k+1}, \quad k\in \mathcal{N}.
    \end{align}
\end{subequations}
The weights are $Q_{c,xx}=1.0$ and $Q_{c,uu}= \text{diag}([1.0, 1.0])$, the sampling time $T_s = 1.0$, and initial state vector is $\bs{x}_0 = \begin{bmatrix} 0.0 & 1.0 \end{bmatrix}^T$ with the covariance $P_0=\text{diag}([0.1,0.1])$.

In this section, we test and compare the proposed numerical methods and investigate the distribution of the stochastic costs via Monte Carlo simulations. 
\subsection{Discretization of LQ-OCP}
\begin{figure}[tb]
    \centering
    \includegraphics[width=0.49\textwidth]{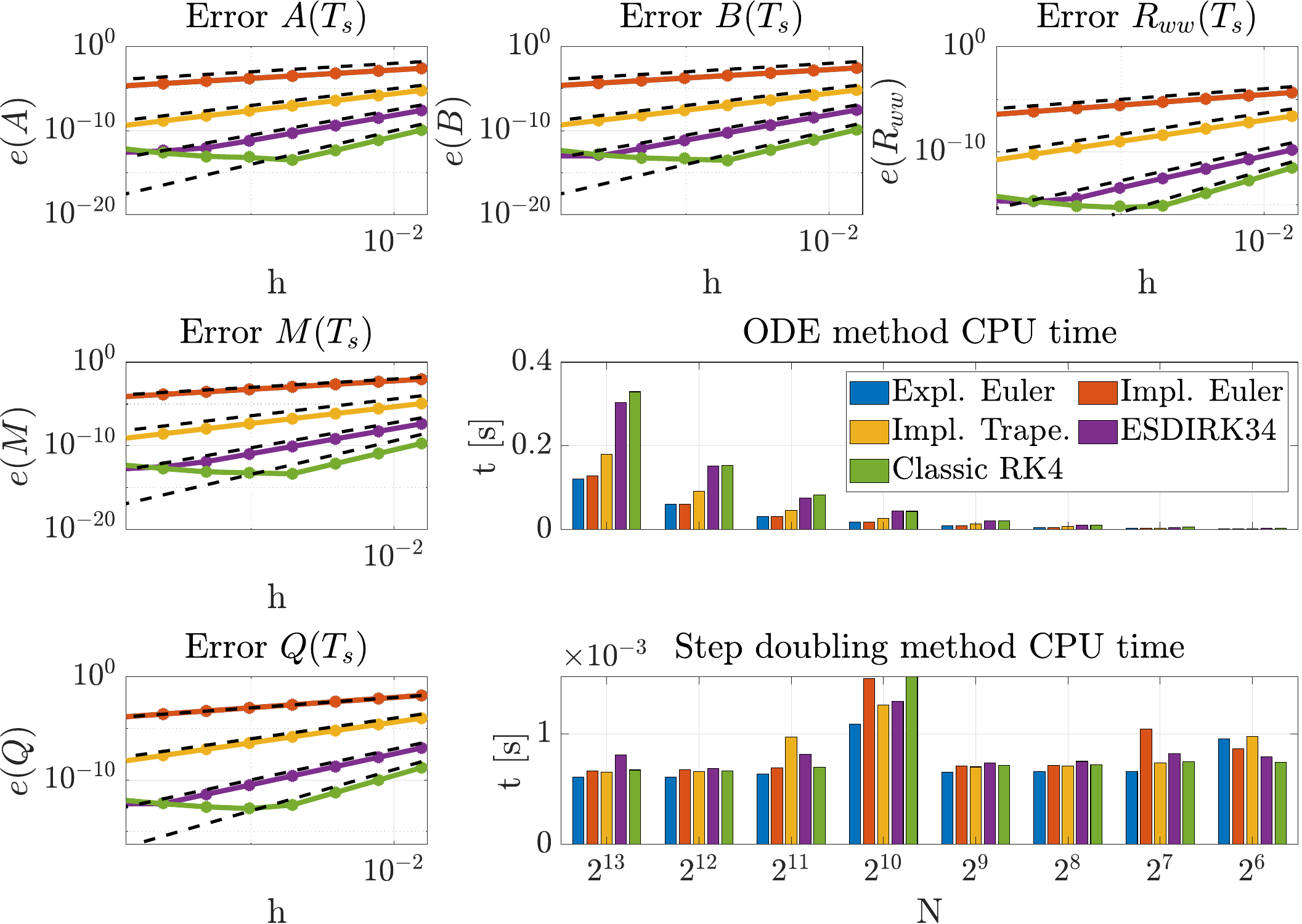}
    \caption{The error and CPU time of the ODE methods and the step-doubling methods with different discretization methods. The error is $e(i) = |i(T_s) - i(N)|$ for $ i \in [A, B, R_{ww}, M, Q]$, where 
    $i(T_s)$ is the true result from the matrix exponential method.}
    \label{fig:ErrorPlotsandCPUTime_LQDiscretization}
\end{figure}
Fig. \ref{fig:ErrorPlotsandCPUTime_LQDiscretization} describes the error and CPU time of ODE and step-doubling methods. The true solution of ($A,B,R_{ww},Q,M$) is calculated using the matrix exponential method. The results of the step-doubling method (dot plots) have the same error as the results of the ODE method (line plots). All methods have the correct convergence order (indicated by dashed lines). In bar plots, the CPU time of the ODE method increases as the integration steps and the stages of the discretization method increase. However, the CPU time of the step-doubling method is stable at around 0.6 ms.  
\begin{table}[tb]
\centering
    \caption{CPU time and error of the scenario using classic RK4 with $N=2^8$}%
    \label{tab:NumericalExperiment-ErrorsandCPUTime}
    \begin{tabular}{ccccc}
    \hline
                & Unit & Matrix Exp. & \multicolumn{1}{l}{ODE Method} & \multicolumn{1}{l}{Step-doubling} \\ \hline
    $e(A)$      & [-] & -           & $7.49 \cdot 10^{-12}$          & $7.49 \cdot 10^{-12}$             \\
    $e(B)$      & [-] & -           & $8.33 \cdot 10^{-12}$          & $8.33 \cdot 10^{-12}$             \\
    $e(R_{ww})$ & [-] & -           & $9.73 \cdot 10^{-11}$          & $9.73 \cdot 10^{-11}$             \\
    $e(M)$      & [-] & -           & $1.25 \cdot 10^{-11}$          & $1.25 \cdot 10^{-11}$             \\
    $e(Q)$      & [-] & -           & $2.03 \cdot 10^{-13}$          & $2.03 \cdot 10^{-13}$             \\
    CPU Time    & [ms] & 0.74     & 9.5                         & 0.68      \\ \hline
    \end{tabular}
\end{table}

Table \ref{tab:NumericalExperiment-ErrorsandCPUTime} describes the error and CPU time of ODE and step-doubling methods with the classic RK4 method applied with the integration step $N=2^8$. The step-doubling method has the same error as the ODE method. The ODE method is the slowest among the three methods and takes 9.5 ms, while the matrix exponential and the step-doubling methods spend 0.74 ms and 0.68 ms, respectively. 
\subsection{Distribution of stochastic LQ-OCP}
\begin{figure}[tb]
    \centering
    \includegraphics[width=0.5\textwidth]{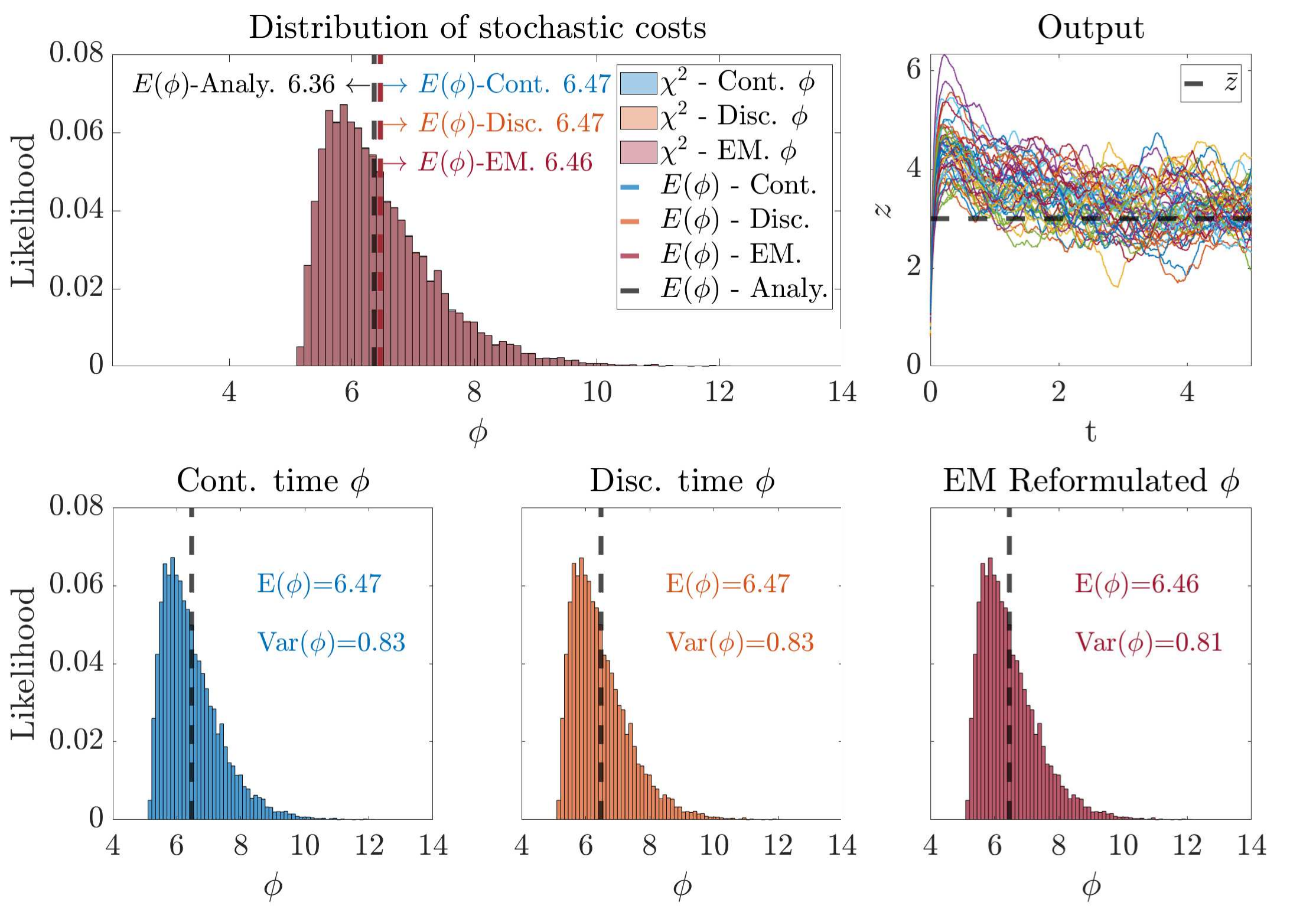}
    \caption{The likelihood of the cost functions of continuous-time and discrete-time stochastic LQ-OCPs with 30000 Monte Carlo simulations. The Cont., Disc. EM. indicate the continuous-time, discrete-time, and EM reformulated stochastic costs, respectively. $E\{\phi\}$-Analy. is the analytic expectation described by \eqref{eq:analyticMeanOfStochasticCosts}, where the continuous-time element $\text{tr}(Q_{c,ww}P_w)$ is solved using the EM method with $N=2^8$.}
    \label{fig:IntrodistributionPlots}
\end{figure}
Fig. \ref{fig:IntrodistributionPlots} shows the likelihood of the continuous-time stochastic costs \eqref{eq:Cont-Stc-LQ-OCP}, discrete stochastic costs \eqref{eq:Dist-Stc-LQ-OCP}, and EM reformulated stochastic costs \eqref{eq:EMReformulatedQP} via 30,000 Monte Carlo simulations. We apply the EM method to solve these stochastic cost functions with $N=2^8$ integration steps.

The simulation results show that the continuous-time cost has the same distribution as the discrete and EM reformulated stochastic costs. There is an offset between the analytic expectation ($E(\phi)=6.36$, obtained by~\eqref{eq:analyticMeanOfStochasticCosts}) and the other numerical expectations (6.47 for continuous- and discrete-time cases and 6.46 for the EM case). We consider it reasonable to have numerical errors since we cannot take the limit of $N \rightarrow \infty$ in experiments. 
% The estimated cost distribution is generated using the MATLAB toolbox "Generalized chi-square distribution"~\cite{Das2021genchi2} based on Proposition \ref{prop:DistributionofStochasticCost}, and it is identical to both the continuous-time and discrete-time cases. 

%% file: tex/Conclusion.tex
\section{Conclusions}
\label{sec:Conclusion}
In this paper, we have discussed the discretization of both deterministic and stochastic LQ-OCPs and proposed three numerical methods. In our propositions, LQ discretization is converted into explicit, neat differential equation systems. We further extend the problem from the deterministic to the stochastic case and illustrate the stochastic cost adheres to a generalized $\chi^2$ distribution. The proposed numerical methods are tested and compared in the numerical experiment, and its results indicate: 1) the step-doubling method is the fastest among the three methods while retaining the same accuracy and convergence order as the ODE method, 2) the discrete-time LQ-OCP derived by the proposed numerical methods is equivalent to the original problem in both the deterministic and stochastic cases. 

% Further, the cost of stochastic LQ-OCP adheres to a generalized $\chi^2$ distribution

% We have studied the discretization of LQ-OCP, and proposed three numerical methods for LQ discretization. Our methodologies ensure that the continuous-time LQ-OCP can be discretized by solving systems of differential equations $(A,B,R_{ww},Q,M)$, and the cost of stochastic LQ-OCP adheres to a generalized $\chi^2$ distribution. The numerical experiment tests and compares the proposed numerical methods, its result shows: 1) the step-doubling method is the fastest among all three numerical methods while retaining the same accuracy and convergence order as the ODE method, 2) the discrete-time LQ-OCP derived by the proposed numerical methods has the same distribution as the continuous-time LQ-OCP. It will be interesting to investigate the discretization of LQ-OCP with time delays, which is essential in practical applications of optimal control and estimation techniques.

%% file: tex/AppendixProofs.tex
\section*{Distribution of stochastic costs} \label{app:DistributionProofs}
To evaluate the distribution of the costs, consider the EM discretization of the stochastic system with a fine time step $\delta t = \frac{T_s}{n}$, $t_i=i \delta t$ for $i = 1,\ldots, n$ is 
\begin{subequations}
    \begin{equation}
        x_{k,i} = \overbrace{(I+\delta t A_c)^i}^{A_i} x_k + \overbrace{ \left(\sum_{j=0}^{i-1} A(j) \delta t B_c \right) }^{B_i} u_k + G_i \bs{w}_{k},
    \end{equation}
where 
\begin{alignat}{3}
    G_i &= \begin{bmatrix}
        A_{i-1}G_c & A_{i-2}G_c & \ldots & 0_{n-i}
    \end{bmatrix},
    \\
    \bs{w}_k &= \begin{bmatrix}
        \Delta \bs{w}_{k,1}' & \Delta \bs{w}_{k,2}' & \ldots & \Delta \bs{w}_{k,n}'
    \end{bmatrix}'.
\end{alignat}
\end{subequations}
The EM expression for the extended state vector $[x_k; u_k]$ is 
\begin{subequations}
    \begin{equation}
        \begin{bmatrix}
            \bs{x}_k \\ u_k
        \end{bmatrix} = \begin{bmatrix}
            A_n^k \\ 0
        \end{bmatrix} \bs{x}_0 +\begin{bmatrix}
            \Theta_{u,k} \\ I_{u,k}
        \end{bmatrix} U_{\mathcal{N}} + \begin{bmatrix}
            \Theta_{w,k} \\ 0
        \end{bmatrix} \bs{W}_{\mathcal{N}},
    \label{eq:EMSystemDiscretization}
    \end{equation} 
where $\bs{W}_{\mathcal{N}}$ and $U_{\mathcal{N}}$ are vectors of the random $\bs{w}_k= I_{w,k} \bs{W}_{\mathcal{N}}$ and the input $u_{k}=I_{u,k} U_{\mathcal{N}}$ over the horizon $\mathcal{N}$, and 
    \begin{alignat}{3}
        \Theta_{u,k} &= \begin{bmatrix}
            A^{k-1}_nB_n & A^{k-2}_nB_n & \ldots & 0_{\mathcal{N}-k}
        \end{bmatrix},
        \\
        \Theta_{w,k} &= \begin{bmatrix}
            A^{k-1}_n G_n & A^{k-2}_n G_n & \ldots & 0_{\mathcal{N}-k}  
        \end{bmatrix}.
    \end{alignat}
\end{subequations}
The corresponding EM expression of the discrete stochastic cost function $\phi$ is 
\begin{equation}
    \begin{split}
        \phi &= \sum_{k \in \mathcal{N}} l_k( \bs{x}_k, u_k) +  l_{s,k}(\bs{x}_k, u_k)
        \\
        &= \sum_{k \in \mathcal{N}} \frac{1}{2} \begin{bmatrix} \bs{x}_k \\ u_k \end{bmatrix}' Q \begin{bmatrix} \bs{x}_k \\ u_k
        \end{bmatrix} + q_k' \begin{bmatrix} \bs{x}_k \\ u_k
        \end{bmatrix}  + \rho_k, \label{eq:EMstochcost}
    \end{split}
\end{equation}
where 
\begin{subequations}
    \begin{alignat}{3}
        q_k &= M \bar{z} _k + \Omega \bs{w}_k,
        \\
        \begin{split}
            \rho_{k} &=\sum_{i=1}^n \frac{1}{2} \bs{w}_k' (G_i' \bar{Q}_{c,ww} G_i) \bs{w}_k + (- \delta t G_i' \bar{Q}_c\bar{z}_k)' \bs{w}_k \\
            & \qquad + \frac{1}{2} \bar{z}_k' \bar{Q}_{c} \bar{z}_k.
        \end{split}
        \\
        \Gamma_i &= \begin{bmatrix} C_c & D_c \end{bmatrix} \begin{bmatrix} A_i & B_i \\ 0 & I \end{bmatrix},
        \\
        \Omega &= \sum_{i=1}^n \delta t \Gamma_i' Q_{c} \begin{bmatrix} C_c' & 0 \end{bmatrix}' G_i ,
    \end{alignat}
\end{subequations}
Using \eqref{eq:EMSystemDiscretization} to substitute $[x_k;u_k]$, we reformulate the quadratic problem into isolated stochastic form \eqref{eq:EMstochcostN} with deterministic weights $Q_{\mathcal{N}}$, $q_{\mathcal{N}}$ and $\rho_{\mathcal{N}}$. 
\begin{subequations}
    \begin{align}
    \phi = \frac{1}{2} \begin{bmatrix} \bs{x}_0 \\ \bs{W}_{\mathcal{N}} \end{bmatrix}' Q_{\mathcal{N}} \begin{bmatrix} \bs{x}_0 \\ \bs{W}_{\mathcal{N}} \end{bmatrix} + q_{\mathcal{N}}' \begin{bmatrix} \bs{x}_0 \\ \bs{W}_{\mathcal{N}} \end{bmatrix} + \rho_{\mathcal{N}}, \label{eq:EMstochcostN}
    \\
    \begin{bmatrix}
      \bs{x}_0 \\ \bs{W}_{\mathcal{N}} 
    \end{bmatrix}
    \sim N(\bar{m}, \bar{P}), \quad \bar{m} = \begin{bmatrix}
        x_0 \\ 0
    \end{bmatrix},
    \quad 
    \bar{P} = \begin{bmatrix}
    P_0 & 0 \\ 0 & P_w
    \end{bmatrix},
    \end{align}
\end{subequations}
where the state vector of the reformulated cost function is normally distributed, and $\bs{W}_{\mathcal{N}}$ has the covariance $P_w = \text{diag}(I\delta t ,I \delta t, \ldots, I \delta t)$.

Based on the theory of integrating the normal in the quadratic domain~\cite{Das2021genchi2}, the stochastic cost $\phi$ follows a generalized $\chi^2$ distribution. The quadratic-form of its expectation and variance can be computed as \eqref{eq:meanOfStochasticCosts} and  \eqref{eq:varOfStochasticCosts}.
% shown to be % (reversed argumention of \cite{Das2021genchi2})
% \begin{subequations}\label{eq:StochasticPar}
%     \begin{alignat}{3}
%         & E\set{\phi} = \frac{1}{2}\bar{m}'Q_{\mathcal{N}} \bar m + q_{\mathcal{N}}'\bar m + \rho_{\mathcal{N}} + \frac{1}{2}\text{tr}\left( Q_{\mathcal{N}} \bar P\right)
%         \label{eq:meanOfStochasticCosts} 
%         \\ 
%         \begin{split}
%         & V\set{\phi} = q_{\mathcal{N}}' \bar P q_{\mathcal{N}} + 2m'Q_{\mathcal{N}} \bar P q_{\mathcal{N}} + \bar{m}'Q_{\mathcal{N}} \bar P Q_{\mathcal{N}} \bar m \\
%         & \qquad \qquad + \frac{1}{2}\text{tr}(Q_{\mathcal{N}} \bar P Q_{\mathcal{N}} \bar P) 
%         \label{eq:varOfStochasticCosts}
%         \end{split}
%     \end{alignat}
% \end{subequations}

The original cost \eqref{eq:Cont-Stc-LQ-OCP} is equivalent to the discrete cost when taking the limit
% As the system is linear, the integrand in the original cost \eqref{eq:Cont-Stc-LQ-OCP} is quadratic, and the stochastic $d\mathbf{\omega}(t)$ is continuous, it can be shown that the Riemann integral exist:
\begin{align}
    \int^{t_{0}+T}_{t_0}l_c(\Tilde{z}(t))dt
     = \lim_{n\rightarrow\infty}\sum^n_{j=1}l_c(\Tilde{z}(t_0+j\delta t))\delta t, \quad \delta t= \frac{T}{n}.
\end{align}
Thus, the stochastic cost $\phi$ of the continuous-time LQ-OCP is a generalized $\chi^2$-distribution variable, with mean and variance taken in the limit of \eqref{eq:meanAndVariance}.